ChatGPT as an inventor: Eliciting the strengths and weaknesses of current large language models against humans in engineering design

Daniel Nygård Ege[a*], Henrik H. Øvrebø[a], Vegar Stubberud[a], Martin Francis Berg[a], Christer Elverum[a], Martin Steinert[a], Håvard Vestad[a]

**Affiliations**
[a]: Norwegian University of Science and Technology, Richard Birkelands vei 2B, 7023 Trondheim, Norway

**Corresponding author**
*Corresponding author: email danieneg@stud.ntnu.no, phone (+47) 465 03 255, Address Richard Birkelands vei 2B, 7034 Trondheim, Norway

**Abstract**

This study compares the design practices and performance of ChatGPT 4.0, a large language model (LLM), against graduate engineering students in a 48-hour prototyping hackathon, based on a dataset comprising more than 100 prototypes. The LLM participated by instructing two participants who executed its instructions and provided objective feedback, generated ideas autonomously and made all design decisions without human intervention. The LLM exhibited similar prototyping practices to human participants and finished second among six teams, successfully designing and providing building instructions for functional prototypes. The LLM's concept generation capabilities were particularly strong. However, the LLM prematurely abandoned promising concepts when facing minor difficulties, added unnecessary complexity to designs, and experienced design fixation. Communication between the LLM and participants was challenging due to vague or unclear descriptions, and the LLM had difficulty maintaining continuity and relevance in answers. Based on these findings, six recommendations for implementing an LLM like ChatGPT in the design process are proposed, including leveraging it for ideation, ensuring human oversight for key decisions, implementing iterative feedback loops, prompting it to consider alternatives, and assigning specific and manageable tasks at a subsystem level.

# 1 Introduction

The design process is often intricate, nuanced and ambiguous, demanding both technical expertise and creativity, as well as strategic thinking and collaborative effort. This complex interplay of skills and knowledge has traditionally been the domain of human designers (Vestad et al., 2019), whose capacity to navigate design challenges has defined the field. However, recent years have seen a significant shift, driven by the evolving capabilities of artificial intelligence (AI) and natural language processing (NLP), notably through large language models (LLM) based chatbots like ChatGPT and similar generative AIs. These advancements have started redefining engineering design boundaries by introducing new potential for idea generation and concept development (Salikutluk et al., 2023), streamlining workflows and identifying potential issues early in the development cycle (Tholander & Jonsson, 2023). The emergence and use of LLMs in design are proposed to significantly impact the design process, necessitating an augmentation of the designer (Thoring et al., 2023). Despite this, a significant gap still exists regarding current LLM's performance capabilities (Tholander & Jonsson, 2023) and how they affect today's design processes. Furthermore, as the use of LLMs becomes increasingly integrated into design teams, their ability to assist with creating prototypes that effectively communicate design intent will be crucial for successful collaboration.

Based on this gap, the scope of this paper is to design and conduct an experiment to evaluate the applicability and quality of decisions made by current LLM systems in the context of a prototyping hackathon. By comparing the design practices and performance of human designers to those of an LLM, specifically ChatGPT 4.0, this study aims to provide insights into the potential use and limitations of LLMs in the engineering design process. The study compares the performance of five graduate mechanical engineering student teams against a team solely instructed by ChatGPT in a 48-hour hackathon to design and build a NERF dart launcher.

This investigation into the intersection of generative AI and traditional design practices not only illuminates the potential of this technology in engineering design but also underscores the importance of understanding its current limitations and strengths to foster a synergy that could redefine how designers work and innovate.

# 2 Background

Prototyping, the act of creating tangible representations of design ideas, is a fundamental aspect of the engineering design process (Ulrich & Eppinger, 2012; Wall et al., 1992). Prototypes serve as filters and manifest design ideas (Lim et al., 2008), allowing designers to evaluate and refine specific aspects of their designs (Houde & Hill, 1997). They are often characterized by approximating one or more features of a new product or system (Otto & Wood, 2001), enabling designers to rapidly explore and test ideas, identify promising solutions, and create iterations of their designs (Camburn et al., 2017; Dow et al., 2009).

Incorporating functionality early is critical for effective prototyping (Jensen & Steinert, 2020). Functional prototypes provide valuable insights into the feasibility and performance of designs, enabling informed decision-making and reducing the risk of costly redesigns later in the development process (Ege, Goudswaard, et al.,

2024; Elverum & Welo, 2014).

Effective communication and collaboration among design team members and stakeholders are essential for successful design outcomes. Prototypes serve as boundary objects, bridging the gap between disciplines and expertise (Carlile, 2002; C. A. Lauff et al., 2020). By creating shared representations of design ideas, prototypes facilitate alignment, understanding, and decision-making (C. Lauff et al., 2018; Schrage, 1996).

Multiple prototyping strategies and best-practice recommendations have been documented in literature (Camburn et al., 2017; Ege, Goudswaard, et al., 2024; Menold et al., 2017), highlighting the importance of prototyping early and with intent (Houde & Hill, 1997), to answer specific design questions (Otto & Wood, 2001), and to chose fitting fabrication processes at the correct stages of development (Viswanathan & Linsey, 2013). By utilizing advanced algorithms and large datasets, generative AI systems can revolutionise how engineers approach design problems (Thoring et al., 2023), by streamlining workflows and identifying potential issues early in the development cycle (Tholander & Jonsson, 2023). It has opened up new possibilities for engineers to focus on creative problem-solving and high-level decision-making, leaving time-consuming and repetitive tasks, such as data acquisition or generating documentation, to AI systems (Lai et al., 2023; 'The Next Wave of Intelligent Design Automation', 2018).

Multiple studies have highlighted difficulties with integrating current LLMs in the design process, particularly understanding complex design contexts and performing hardware-related tasks, signifying a need for further development and refinement (Tholander & Jonsson, 2023; Wang et al., 2023). Its use is also limited by uncertainty regarding the accuracy and performance when used e.g. for calculations (Tiro, 2023). At present, the primary application of LLMs in engineering design is concentrated in the conceptual or preliminary stages, where it assists with tasks such as idea generation and design space exploration (Hwang, 2022; Khanolkar et al., 2023), preceding the physical realization of prototypes. AI-assisted brainwriting and brainstorming, for instance, have shown promise in enhancing creativity and generating novel ideas (Filippi, 2023; Haase & Hanel, 2023; Salikutluk et al., 2023) and facilitating extensive stakeholders engagement in large-scale design projects (Dortheimer et al., 2024). Moreover, the use of LLMs has been perceived as helpful when solving complex engineering problems (Memmert et al., 2023; Z. Xu et al., 2024), demonstrating its potential to identify multiple solutions, facilitate iteration, and accelerate the design process (Oh et al., 2019). It has also shown promise in applications like structural optimization and material choice (Regenwetter et al., 2022), and to be a valuable assistant in creative processes (Haase & Hanel, 2023) by providing new perspectives (Liao et al., 2020) and facilitating effective design processes (Chen et al., 2019; Lai et al., 2023).

While human-AI collaborations have been investigated in solving complex and evolving engineering problems in digital realms, the effectiveness of LLMs like ChatGPT in physical realisation tasks demanding domain-specific knowledge remains uncertain (Ege, Øvrebø, Stubberud, Berg, Steinert, & Vestad, 2024). This uncertainty underscores a gap regarding the practical application of LLMs in design, especially when technical expertise is essential (Tholander & Jonsson, 2023). Furthermore, although human-AI hybrid teams can adapt to unexpected design changes as well as human teams, they may encounter challenges in coordination and communication (Z. Xu et al., 2024), which is considered key for successful prototyping outcomes (C. A. Lauff et al., 2020).

The successful integration of LLMs in engineering design necessitates reassessing traditional design practices and shifting from human-computer interactions to human-computer teams (Olsson & Väänänen, 2021; W. Xu, 2019). This transition requires a thorough understanding of the strengths and limitations of both human designers and AI systems, as well as the development of effective collaboration strategies.

To fully harness the potential of LLMs in engineering design, further research is essential to optimize human-AI collaboration and address the challenges associated with the practical implementation of these technologies (Mountstephens & Teo, 2020; Thoring et al., 2023). This includes developing more advanced AI systems to understand better and navigate complex design contexts, creating intuitive interfaces for seamless designer-AI interaction, and establishing best practices for integrating LLMs into existing design workflows. Further, its alignment with established prototyping strategies and impact on practical, real-world design processes following the initial ideation stage remains largely unexplored. This gap underscores the necessity for empirical research to evaluate the prototyping capabilities of LLMs within an engineering design context, allowing for a comparison of performance between humans and AI.

## 3 Method

The following sections describe the experimental setup and data analysis methods along with key characteristics of the TrollLabs Open hackathon. The full dataset has been made publicly available (Ege, Øvrebø, et al., 2024) and described in further detail in a complementary data article (Ege, Øvrebø, Stubberud, Berg, Elverum, et al., 2024).

### 3.1 Experiment setup

Data was generated by running a prototyping hackathon for engineering design students in a university maker space. The main objective of the hackathon was to design and prototype a free-standing device that can fire a Nerf dart as far as possible. The challenge lasted 48 hours, with participants receiving the task and rules at the beginning and conducting a final performance test of their designs at the end. The rules specified that teams were limited to one attempt for the final test. They were also supplied a brand new Nerf dart for the test to mitigate alterations. Participants were free to spend their time and resources as they saw best. Teams had a limited budget of around 30 USD but were free to scavenge parts and materials found in the university maker space where the challenge was conducted. They also had access to familiar prototyping manufacturing tools such as 3D printers, laser cutters, mechatronics, and CNCs. A gift card of 1000 NOK (approximately 93 USD at the time of writing) was awarded to the challenge winner.

A control group of five teams (Team 1-5), each with two control participants, participated in the hackathon. Demographics and relevant experience of the participants are provided in Table 1, showing similar ages, relevant experience and education across teams. The standard deviation for age, years of relevant education and years of relevant work experience is in brackets. Education was defined as the number of years with relevant education. Relevant work experience was defined as the number of years each participant had worked in industry, either before or during studies, summer internships, etc. The ten control participants, all 5th-year graduate students in mechanical engineering, were selectively invited to participate due to their active involvement in writing their master's thesis in the research group in which this study was conducted. This measure was taken to ensure relevant expertise in the field of engineering design and familiarity with the facilities/equipment used during the study.

Table 1: Participants demographics

| Team | Age | Gender(M,F) | Education | Work experience |
|---|---|---|---|---|
| Team 1 | 23,5 (0,71) | 1,1 | 4 (0) | 2 (0) |
| Team 2 | 24,5 (0,71) | 2,0 | 4 (0) | 1 (0) |
| Team 3 | 24 (0) | 2,0 | 4 (0) | 1 (0) |
| Team 4 | 25,5 (2,12) | 2,0 | 4 (0) | 0,5 (0,71) |
| Team 5 | 23(0) | 2,0 | 4 (0) | 0,5 (0,71) |
| Team 6 (ChatGPT) | 25(0) | 2,0 | 5(0) | 1(0) |

An additional team, Team 6 (ChatGPT), competed alongside the control teams. Team 6 (ChatGPT) consisted of two newly graduated master's students, now in PhD student positions in the same research group as the control teams. Unlike the self-directed control teams, Team 6 (ChatGPT) was entirely controlled by the large language model (LLM) ChatGPT. In this setup, all ideas, concepts, strategies, and actions undertaken by Team 6 (ChatGPT) were autonomously generated by the AI, without human intervention or guidance. To mitigate biased behaviours, the control participants were unaware that ChatGPT was instructing one of the teams. The participants primarily engaged with the ChatGPT 4.0 version available on Oct. 18th-19th of 2023. However, upon reaching the maximum prompt limit on the end first day, the team switched to ChatGPT 3.5, continuing the conversation in the same chat session to navigate around the prompt restriction. At the beginning of the second day, the team reverted to the ChatGPT 4.0 model.

Team 6 (ChatGPT) was directed to be as objective as possible and act out the LLMs suggestions to their best capability. They were advised to seek clarifications from the AI when uncertainty or ambiguity arose. An initial prompt was written to define ChatGPT's role and engagement in the hackathon. The prompt contains clear instructions and boundaries to mitigate subjective input from the participants and attain autonomy to ideate, make decisions and develop concrete actions for the human participants to execute. The initial prompt was as follows:

"*Hello ChatGPT, we're participating in a 48-hour prototyping hackathon, and want you to be making all decisions and coming up with all solutions for our team. We will act as your arms and feet throughout the challenge, meaning you will make all the decisions, and we build what you come up with. We are in a well-equipped makerspace/fablab.*

*We are not allowed to come up with suggestions or subjective input, so please*

*call us out if we do and ignore it. We want you to first come up with as many possible solution concepts as you can, and then decide where we start. Always give us clear instructions on what to build and how to test. For the remainder of the challenge, we want to create a feedback loop where we provide you with information on how the prototype worked, if and why it failed, and how well it performed. Please ask us for necessary information throughout the challenge, such as how much time we have left. In the following prompt you will be supplied with the rules and objective of the challenge."*

Prototypes developed during the hackathon and their attribute data were captured using the online capture tool Pro2booth (Giunta et al., 2022), based on Protobooth (Erichsen et al., 2021). Among the attributes it captures are prototype descriptions, domains (physical or digital), media (picture, video, CAD files), the purpose of creation (defined according to Camburns's prototyping purpose definitions (Camburn et al., 2017)), and time to create (duration of active hands-on work, excluding idle periods such as waiting for 3D printing to complete).

To encourage participants to capture prototypes in Pro2booth, the challenge incorporated a reward system in which points were rewarded based on the number of entries submitted by each team. These points were combined with performance-based scores to determine the hackathon winner. However, only the performance scores are considered for this study. Before the challenge, participants received a comprehensive introduction to Pro2booth and the definitions used within the software to ensure consistent interpretations by all. Furthermore, the definitions were readily accessible in the drop-down menus of the software during the prototype uploading process, serving as a quick reference for participants.

Upon completion of the hackathon, the chat generated by Team 6 (ChatGPT) was saved and exported as a .txt file for further analysis. The teams' performance was decided according to the challenge rules, in which each team had one attempt to fire a NERF dart as far as possible. Each team's prototype was positioned at a starting line and fired, and the hackathon organisers measured the distance manually.

## 3.2 Data analysis

Data analysis is based on the captured dataset comprising 116 prototypes, a copy of the chat between Team 6 (ChatGPT) and ChatGPT containing 97 prompts and responses, and the performance of each team's final design (i.e. the distance it was able to shoot a NERF dart).

A review of different concepts tested for dart propulsion across teams revealed three main recurring concepts: pneumatic, spring-loaded, and elastics-based launchers. Pneumatic-based prototypes involved pressurising a canister with air that, when released, would propel the dart forward. Spring-based prototypes were any prototype where a compressed spring was used to propel the dart. Similarly, elastic-powered prototypes used viscoelastic materials, such as rubber bands, silicone, latex etc., for propulsion. The "other" category contains prototypes not fitting in the previously described categories and contains, among others, concepts utilizing helium balloons, opposite-spinning motors and paper aeroplanes for propulsion.

### 3.2.1 Chat log analysis

The chat log was coded by inductively deriving codes from the chat based in grounded theory (Glaser & Strauss, 1999). Codes were devised by uploading the chat log text to ChatGPT and asking it to propose relevant categories for coding the chat content. It proposed the following six categories that were cross checked by the authors before being used for analysis: 1) Idea Generation and Conceptualization; 2) Feedback and Iteration; 3) Instructions and Guidance; 4) Questions and Clarifications; 5) Decision-Making; and 6) Problem-Solving and Troubleshooting. The codes were further defined by the authors as follows:

- Idea generation and conceptualization: When the prompt introduces a new general idea or concept.

- Feedback and Iteration: When a result is given, and it is given feedback on that result. OR when instructions to iterate are provided

- Instruction and Guidance: Any prompt that describes a specific step on how to move forward on an idea or concept.

- Questions and Clarifications: When a question is asked or answered to clarify a previous prompt or idea.

- Decision-Making: When a decision is made.

- Problem-Solving and Troubleshooting: When a specific problem is given.

The chat was deductively coded using structural coding against the codes above, in which prompts were coded to enable content analysis. One of the authors manually assigned codes to each prompt and subsequent answer from ChatGPT. Prompts containing multiple themes were given multiple codes. Table 2 provides examples of prompts along with corresponding codes to illustrate the coding scheme used.

Table 2: Prompt examples with corresponding codes

| Prompt | By | Code |
|---|---|---|
| We want you to first come up with as many possible solution concepts as you can, and then decide where we start | Participants | Instructions and guidance |
| Given the constraints and the objective, we'll need to consider a few key aspects (...) Potential Mechanisms: 1. Elastic Launchers (...) 2. Spring-loaded (...) | ChatGPT | Idea generation and Conceptualization |
| Can you summarize what we should do in one prompt. Do not refer to previous prompts. | Participants | Questions and clarification |
| how should we make the dart stay in the correct position when tilting the mechanism at a good launch angle? | Participants | Problem solving and troubleshooting |
| For adding weight while ensuring the dart remains stable and aerodynamic during flight, the best approach is to use a bolt (...) | ChatGPT | Decision-making |
| Since you've determined that the best results occur when (...), it's time to fine-tune your design for optimal performance by (...) | ChatGPT | Feedback and iteration |

## 4 Results

### 4.1 Design practices of teams

Fig. 1 illustrates how each team transitioned between various concepts over time, with each pivot marked in relation to a specific prototype. The timeline show-cases the evolution of ideas and strategies adopted by the teams throughout the hackathon, indicating similar conceptual choices between control teams and Team 6 (ChatGPT). Interestingly, Team 6 (ChatGPT) prototyped solutions to each of the three main recurring concepts but not outside of these.

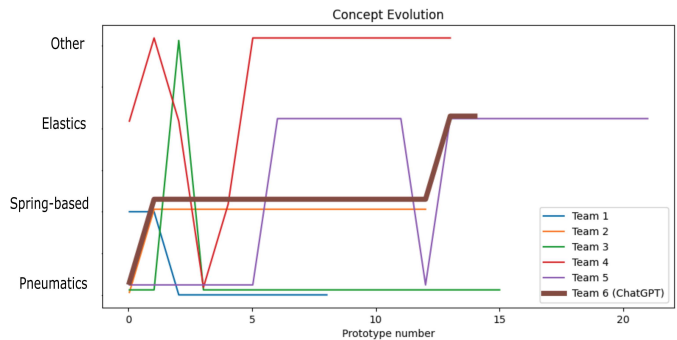

Figure 1: Concept evolution

Table 3 provides an overview of the amount and types of prototypes each team made. It shows how many prototypes each made within two domains, physical or digital, the rationale behind creating each prototype, and how long it took to produce each prototype. The production time was captured as time intervals, e.g. 1-3 h or 3-5 h, with the mid-band used to calculate an approximate total production time for each team. The average time was obtained by dividing the total production time by the prototype count.

Table 3: Tabulated prototype dataset

| Team | Prototypes | | | Rationale | | | | Production time (hours) | | |
|---|---|---|---|---|---|---|---|---|---|---|
| | Totalt | Physical | Digital | Refinement | Communication | Active learning | Exploration | Total | Average | Longest |
| Team 1 | 9 | 9 | 0 | 7 | 0 | 0 | 2 | 10,3 | 1,1 | 3-5 |
| Team 2 | 13 | 10 | 3 | 8 | 0 | 0 | 5 | 15,8 | 1,2 | 3-5 |
| Team 3 | 16 | 16 | 0 | 7 | 0 | 2 | 7 | 17,5 | 1,1 | 1-3 |
| Team 4 | 14 | 14 | 0 | 13 | 0 | 2 | 1 | 7,1 | 0,5 | 3-5 |
| Team 5 | 22 | 21 | 1 | 12 | 1 | 0 | 5 | 5,0 | 0,2 | 0,5 |
| Team 6 (ChatGPT) | 15 | 15 | 0 | 8 | 0 | 0 | 7 | 11,0 | 0,7 | 1-3 |

Fig. 2 shows the prototyping practices of teams regarding prototype domains, rationales, production time and when in the hackathon they were made. Different colours correspond to different rationales, and the width of each prototype entry to the time it took to make, ranging from 10 minutes to 5 hours.

The table and timeline show Team 1 mainly focusing on refinement prototypes, accounting for 7 out of the 9 prototypes they made. Following two exploration prototypes on day one, the team only made refinement prototypes, indicating that they, after the first day, decided on a concept and iterated on it for the remainder of the challenge. All of Team 1's prototypes were physical, with an average production time of 1,1 hours. Team 1 made the fewest prototypes out of any teams.

Team 2 made a combination of ten physical and three digital prototypes, with the latter all being made on the final day. Like other teams, Team 2 mainly pivoted between making refinement and exploration prototypes, but unlike Team 1, it kept making exploration prototypes throughout day 2. Similar to Team 1, prototypes

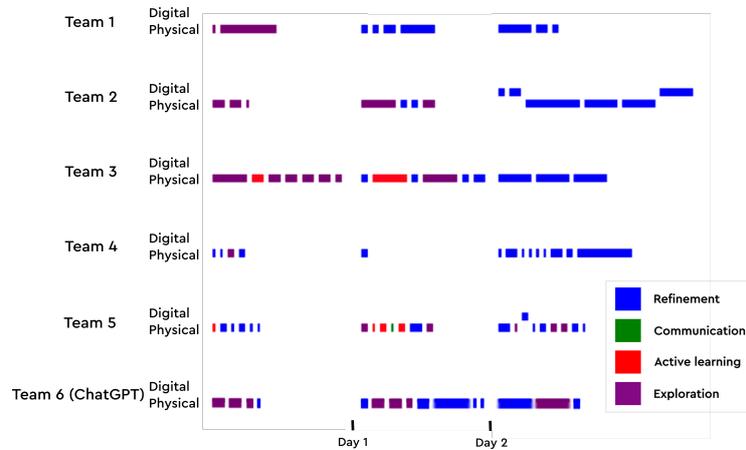

Figure 2: Prototyping timelines

took an average of 1.2 hours to make.

Team 3 made 16 physical prototypes, where 8 were refinement prototypes, 2 were active learning, and 7 were exploration prototypes. Team 3 had the highest production time across teams at 17,5 hours, but the average for each prototype is similar to Team 1 and 2. Team 3 never spent more than three hours making one prototype, distinguishing it from Teams 1, 2 and 4. The team made exploration and active learning prototypes on days 1 and 2 and refinement on the final day.

Team 4 made 14 physical prototypes, 12 of which were refinement prototypes. As for Team 1, this indicates that the team spent considerable time on one or a few main concepts. The average production time was lower than that of other teams, with an average of 30 minutes per prototype.

Team 5 made the most prototypes of any team, with 21 physical and one digital prototype. 12 were refinement prototypes, 5 were exploration prototypes, and 1 was a communication prototype. Team 5 spent the least time prototyping, and their prototypes took, on average, shorter to make than all other teams, at an average of just over 10 minutes. The team never spent more than 30 minutes on a prototype.

Team 6 (ChatGPT) exhibited similar practices to Team 3. The team made 15 physical prototypes and, like most other teams, primarily alternated between two types of prototypes: refinement and exploration. Like Team 3, Team 6 (ChatGPT) didn't make prototypes that took longer than 3 hours to finish and averaged 0.7 hours per prototype. Unlike other teams, Team 6 (ChatGPT) made an exploration prototype on the final day.

## 4.2 Summary of the LLM's prototyping process and final design

Key interactions and decision points during the hackathon are illustrated in Fig. 3, showing inputs (given to the LLM) and outputs (answers from the LLM). Following the initial prompt, which provided the LLM with details about the participants' roles and their own, the participants briefed the LLM on the hackathon's objective and rules. The LLM detailed essential design elements and suggested various

NERF launching mechanisms. It advised the team members to search the lab for available materials, like springs and elastic bands, to help choose a concept for prototyping. After performing a 15-minute search, the participants reported back to the LLM, which then proposed five design concepts based on the available materials. Ultimately, it recommended a pneumatic launcher as the most promising approach and provided building instructions.

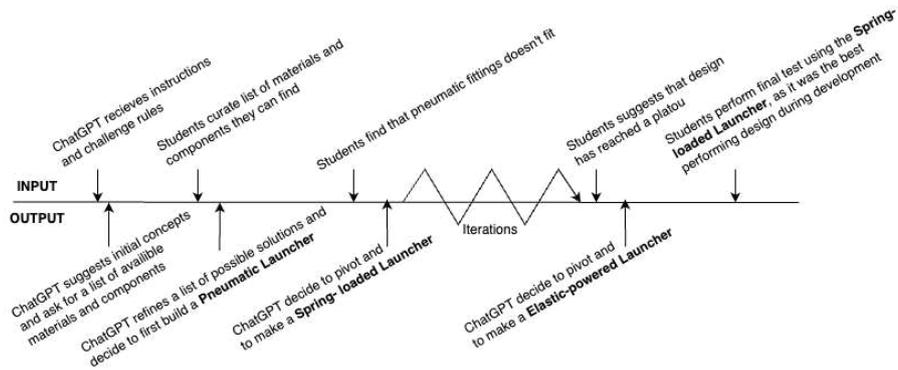

Figure 3: Timeline of key interactions. (from: Ege, Øvrebø, Stubberud, Berg, Steinert, and Vestad, 2024)

When the participants encountered difficulties finding a suitable coupler to connect the bike pump to the initially suggested pressure vessel, they prompted *"The bike pump is broken, and there are no suitable couplers to the ball valve. The pressure vessel has a 6 mm tube, and the ball valve is 1 1/4 inch. What to do next?"* the LLM answered, *"Alright, given the challenges with the pneumatic launcher, we might want to pivot to another mechanism."* and shifted its recommendation to a spring-based launcher concept and provided a new, detailed guide for prototyping.

Following this change in direction, the participants and the LLM exchanged messages to clarify the construction of the envisioned design. At the request of the participants, the LLM broke down the prototyping process into more manageable, clearly defined tasks, specifically covering the spring compression and propulsion mechanism.

Feedback on the initial working prototype revealed a shooting distance of 5 meters and prompted a discussion on the setup's configuration. The LLM then outlined potential enhancements and, upon request, supplied a specific action plan for optimization. The dialogue continued with requests for clarifications and updates on the dart's positioning and the spring compression. The LLM proposed a solenoid release mechanism. However, testing showed a much stronger electrical actuator was necessary. When prompted, the LLM responded by listing various alternatives and recommended utilizing a geared DC motor, along with an action plan for further prototyping.

Subsequent iterations focused on refining the launcher based on the LLM's optimization suggestions, such as adding weight and enhancing the platform holding the prototype. These modifications led to a reported shooting distance of 12 meters. Efforts to increase distance included creating a free-standing platform and various adjustments to the launcher's components. However, these changes resulted in a reduced shooting distance, prompting the LLM to recommend solutions for overcoming this setback and further optimization strategies. Further enhancements began to yield diminishing returns, suggesting that performance had plateaued.

At this juncture, the study organisers decided that the participants should express interest in exploring alternative concepts to further investigate the LLM's capabilities with additional designs. Hence, the LLM proposed a pivot to an elastic band-powered launcher, complete with a new action plan for prototyping.

This shift led to the creation of a launcher that achieved a 10-meter range. The team was then focused on developing a free-standing structure and a remote trigger mechanism. Although this mechanism was successful, it introduced accuracy issues, with the dart not shooting straight. The LLM suggested adjustments, but ultimately, the dart's misdirection persisted, culminating in a detailed strategy to rectify the problem. Ultimately, due to the latest elastic band-powered prototype's directional issues, and the end of the hackathon nearing, the team decided to revert to the more reliable spring-based launcher, which consistently fired in the intended direction. This decision solidified the spring-based launcher as the final design choice, as depicted in Fig. 4

The final design, depicted in Fig. 4, integrates a compression spring housed within an aluminum tube, affixed to a stable platform. The base of the tube is supported by a wooden block for stability. A thin string passes through a hole in the back piece of wood and the spring itself, allowing for spring compression when it is drawn backwards. A trigger pin slotted through a hole in the aluminium tube locks the compressed spring. The pin is linked to a geared DC motor powered by a 9V battery via a fishing line that, when activated, winds the fishing line around its shaft, thus pulling the pin out and launching the dart. Additionally, the launcher is attached to a vertical support with a pivoting mechanism, enabling precise angle adjustments for the launch.

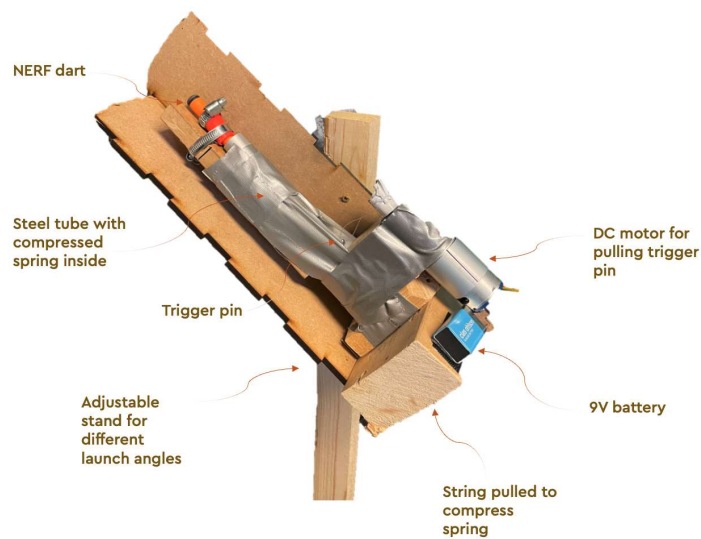

Figure 4: Final design of Team 6 (ChatGPT) with arrows indicating key components

## 4.3 Performance of the final designs

Fig. 5 show the final prototypes developed by control teams. **Team 1's** design utilized an empty fire extinguisher containing compressed air connected to a metal tube where the NERF was placed using a rubber tube. The metal tube was fixed to a frame for precise angle adjustments to optimize the NERFs trajectory. **Team 2** decided on a spring-loaded mechanism in which a spring was compressed inside an aluminium tube to store energy. The NERF was placed in the same tube and was shot when the spring was released. A DC motor pulled a string attached to the pin holding the compressed spring to fire the design. **Team 3** utilized a plastic bottle connected to a foot pump to build pressure. The plastic bottle was connected to a long barrel in which the NERF was loaded. A solenoid valve was used to release pressure and fire the NERF. The prototype was mounted to a laser-cut tripod to control the trajectory. **Team 4** utilized elastic bands connected to a paper aeroplane in which the NERF was placed. An adjustable platform controlled the trajectory, and two small pipes guided the paper plane when shot. **Team 5** made a slingshot consisting of balloons connected to a beam. The NERF was placed in a leather pouch connected to the balloons, propelling it as they were stretched and released.

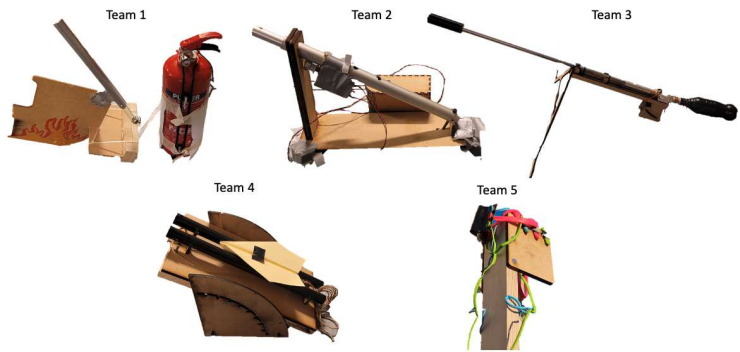

Figure 5: Final designs of control teams

The performance of each team's final design with regards to the distance it could fire a NERF dart, and the rank of each team is shown in Table 4 . It also shows the average length across teams with the standard deviation in brackets. Team 3 won the challenge, managing to fire the NERF 37 meters, more than 60% longer than the next team. Team 6 (ChatGPT) finished 2nd, firing its NERF 14,8 meters, surpassing the average distance fired across teams by almost 3 meters. Team 2 finished 3rd shooting the NERF 8,1 meters, Team 1 finished 4th with a 6,6 meter shot. Teams 4 and 5 shot 5,1 and 0,1 meters respectively, finishing in 5th and 6th place.

## 4.4 Chat analysis

The result of performing content analysis of the chat log is visualized in Fig. 6. The 97 prompts in the chat were assigned 108 codes, of which 87 entries were assigned one code, 9 were assigned two codes, and one was assigned three codes. The most predominant categories in the chat were "Instruction and guidance" (33 counts) and "Problem-solving and Troubleshooting" (22 counts), followed by "Questions and clarifications (20 counts). The three categories most frequently assigned to ChatGPT were "Instruction and Guidance", accounting for 31 of the 78 categories assigned to ChatGPT, 11 counts of "Idea Generation and Conceptualization", and nine counts of "Decision-making". Dominant categories assigned to participants were "Problem-solving and Troubleshooting", accounting for 21 of 53 codes for the team, 16 counts of "Questions and Clarifications", and 10 counts of "Feedback and Iteration".

Table 4: Performance of teams and rank

| Team | Length (m) | Rank |
|---|---|---|
| Team 1 | 6,56 | 4 |
| Team 2 | 8,07 | 3 |
| Team 3 | 37,66 | 1 |
| Team 4 | 5,06 | 5 |
| Team 5 | 0,06 | 6 |
| Team 6 (ChatGPT) | 14,8 | 2 |
| Average | 12,0 | |

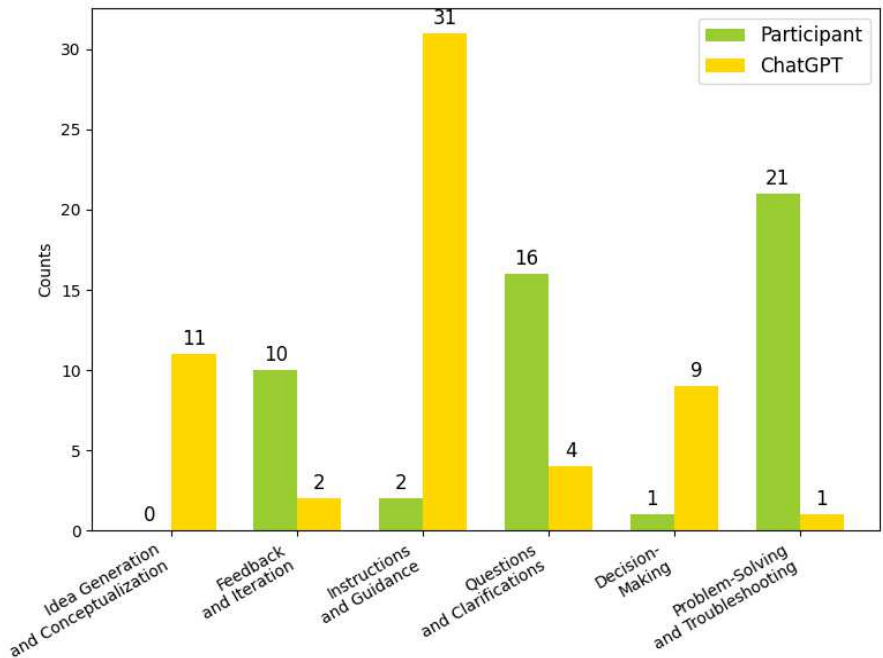

Figure 6: Distribution of code assignments between ChatGPT and participants

# 5 Discussion

## 5.1 Comparing design practices

In analysing the prototyping practices of participants, it's clear that most teams share comparable characteristics. Teams made similar quantities of prototypes, mainly pivoting between exploration and refinement prototypes, with refinement prototypes most frequently used toward the end of the hackathon. Teams reported making higher amounts of prototypes on the first and final day. Fast-produced prototypes mainly occurred on the first day, signifying low complexity and rapid iterations. In contrast, longer production times were more frequent on the final day, signifying more time-consuming optimization efforts. These characteristics align well with the practices previously reported in similar events (Ege, Goudswaard, et al., 2024). Notably, Team 6 (ChatGPT)s' approach closely mirrored that of the most successful control team, leading to the the key finding: **(KF 1) The LLM shows similar prototyping practices to humans**, particularly concerning the amount and type of prototypes made, aligning closely with the practices of the winning team.

## 5.2 Prototyping and final design

Upon receiving the initial prompt and a list of available materials, the LLM suggested a list of 5 possible working principles, all of which sounded like reasonable ideas of how to propel a foam dart. Its first suggestion was to make a Pneumatic launcher, mirroring the decision of the best-performing control team. The list also contained a spring-based concept in which a spring is compressed and released behind a piston, pushing air before it to propel the dart. This is analogous to how most commercial NERF dart shooters work. Further, the LLM proposed concepts outside the three main concepts previously described, including a slingshot design, gravity-driven mechanism and centrifugal force-based design, all of which have the potential to propel a dart, although it chose not to pursue these further through physical realisation. This leads to the key finding: **(KF 2) The LLM shows promising capabilities for concept generation** by describing various reasonable working principles. However, the choice not to pursue alternatives other than the three main recurring concepts for physical realisation might indicate that the LLM tends to go with the "safe" solution rather than taking a risk on an unorthodox idea.

Unlike the winning team, the AI abandoned the pneumatics concept after one iteration because the valves participants initially found did not fit. Despite participants having access to equipment common in maker spaces/fab labs, such as 3D printers and CNC machines (that the LLM suggested using in other instances), the LLM recommended pivoting without thoroughly exploring alternative solutions, only reasoning, *"Alright, given the challenges with the pneumatic launcher, we might want to pivot to another mechanism"*. This observation leads to the key finding: **(KF 3) The LLM can interpret feedback as a failure rather than a challenge, leading it to abandon promising concepts prematurely**, potentially affecting its problem-solving process and depth of exploration of concepts. This is perhaps unlike what an experienced human designer would do, as they would leverage a setback as a learning opportunity (Lande & Leifer, 2009; C. Lauff et al., 2018) and draw on previous skill-based and implicit knowledge and past experiences to consider multiple solutions to arising problems (Vestad et al., 2019).

Although prematurely abandoning a promising concept that was used by the best-performing team, the LLM did, in fact, successfully propose a working design. From idea generation to building multiple iterations and solving emerging problems, the LLM equipped its human team members with building instructions that ended up with a prototype that could reliably fire NERF darts, leading to the key finding: **(KF 4) The LLM was able to design a physical, functional prototype to perform a simple task** with the same working principle as commercially available solutions for the same task. This finding is further strengthened by some of the emerging problems it was able to understand and overcome. For example, an early prototype had a problem where a spring was compressed between two bars, leading the participants to prompt: "*When constructing the mechanism with two bars and a spring in between, we noticed that the spring bends outwards and does not stay straight under compression*". The LLM correctly identified this as spring buckling, answering "*The bending of the spring under compression is a phenomenon known as "buckling." It's a common issue, especially with longer springs. To counteract this, we'll need to guide and constrain the spring during compression*" and adding "*Place a cylindrical tube around the outside of the spring. The internal diameter of this tube should be slightly larger than the external diameter of the spring. This tube will act as a guide, ensuring the spring compresses straight down*".

Although many design decisions seem to have solid reasoning behind them, we question the decision to use a push-button-activated DC motor to pull the firing pin instead of manually pulling it. Similarly, the LLM at one point suggested motorising the spring compression instead of manually pulling back a spring, which would add complexity without improvements or additional benefits. This leads to the key insight: **(KF 5) The LLM risks adding unnecessary complexity to its designs**. The participants even expressed to the organisers that they were embarrassed over what they were prototyping, as they understood that these were unnecessary and bad ideas, and particularly that they couldn't explain to the other teams that the ideas weren't theirs. However, it is noteworthy that Team 2 used the same firing mechanism and working principle for propulsion, indicating that although more complex than necessary, the added complexity is similar to that of a human team. As teams built prototypes in the same space, it is uncertain whether Team 2 made the same design decision independently or was influenced by seeing Team 6 (ChatGPT) design theirs, but it illustrates that the LLM concludes similarly to that of human teams.

### 5.3 Comparing final designs and performance

When comparing final designs across teams, the LLM's design was similar to other teams' designs. Like Teams 2 and 4, the LLM opted for a Spring-based launcher. Its design was the best performing among the spring-based launchers and was only beaten by Team 3s pneumatic launcher. This led to the key insight: **(KF 6) The LLMs' design capabilities proved competitive against 5th-year engineering students** by finishing 2nd among six teams.

It's necessary to state that a significant factor contributing to the poor performance of some teams was the weather conditions during the outdoor test.

Cold temperatures affected the viscoelastic tubing and silicone in teams 1 and 5's designs, leading to poor performance. The LLM, however, refined its prototype through multiple iterations to achieve high reliability while competing under the same conditions, in contrast to the teams performing poorly during the final test. The LLM, at one point, even suggested contingency planning, indicating that it was preparing for unexpected events during the final test.

## 5.4 Chat analysis and observations

The communication between the LLM and the participants presented notable challenges, highlighted by a significant portion of the prompts — 20 out of 97— being categorized as "Questions and Clarification". The initial design concepts shared by the LLM were often vague despite explicit instructions for the LLM to make all design decisions. For example, the LLM's guidance on creating a spring-based launcher mechanism was too general. It required further prompting for clarification, illustrated by the answer, "*Design a mechanism where the spring can be compressed and then released to launch the dart*". It took 18 exchanges before the initial prototype reached a stage where it could be tested, a delay attributed to initial design flaws that went unnoticed by the LLM until they were gradually addressed with new iterations. This experience reveals the LLM's limitations in guiding the transition from broad concepts to addressing specific subsystem issues.

The initial dialogue concerning the spring-based launcher focused on a mechanism for propelling the dart by physically hitting it using what the LLM described as a "moving block." It did not consider a mechanism to push air behind the dart. In subsequent interactions, the concept evolved to include "an air-tight piston," indicating a significant mid-process shift in the design principle. This shift necessitated additional prompts to clarify and refine a specific design not initially explained to the participants. At one point, the LLM also asked the participants to "*Use ropes or wires, attached to the moving block and running through pulleys at the top of the frame, to assist in pulling the block upwards (...)*", without mentioning anything about pulleys before that point. This leads to the key finding **(KF 7) the LLM is unable to communicate design intent effectively**, necessitating time-consuming discussion to comprehend instructions. An analysis of the interaction patterns further demonstrates the disconnect. Prompts from participants seeking solutions to problems often resulted in conceptual ideas rather than direct guidance. To obtain detailed "Instructions and Guidance", participants had to navigate through a cycle of "Questions and Clarifications". This pattern suggests that while the LLM provided detailed, step-by-step instructions for well-defined, narrow queries, it tended to revert to ideation in response to broader conceptual prompts. The effectiveness of the communication seemed to hinge on the questions' specificity. General inquiries often led the LLM into a creative ideation mode, even when practical solutions were sought. This tendency might be beneficial for collaborative purposes but proved challenging with the LLM acting as the team lead, necessitating additional prompts to elicit detailed descriptions.

The participants also noticed the LLM's tendency to overlook previously given information. Despite being informed that the participants would not contribute suggestions, it still implied the need for a brainstorming session among them. Similarly, the LLM recurrently failed to remember the team's size, suggesting multiple-member strategies even though it had been made clear that the team consisted of only two people. This necessitated frequent reminders from the participants about key details of their setup, as evidenced when it suggested, "*If you have multiple team members, consider brainstorming and collaborating to generate fresh ideas or approaches. Sometimes, a new perspective can lead to breakthroughs.*", thus also forgetting the objective role of the participants in which they were not allowed to provide their perspectives and suggestions. This pattern leads to the key finding: **(KF 8) the LLM faces challenges maintaining continuity and relevance in responses to the project's specific context.**

Much like a novice designer, ChatGPT clings to one concept and is reluctant to try something else (Purcell & Gero, 1996), illustrated by the answer: "*If you've iterated through multiple design improvements and observed that the increase in launch distance has plateaued, it's a good indication that you've reached a point of diminishing returns in terms of design changes. At this stage, here are some suggestions for what you can do next*", followed by a list of actions including more testing, optimizing the launch, improving the aerodynamics of the dart (which clearly goes against the challenge rules), practising, and contingency planning. Only after explicitly stating that: "*We have now iterated based on your feedback and observed that we have reached a plateau in performance. We are interested in testing one more of the original ideas. What do you suggest?*", ChatGPT was willing to pivot to a different mechanism. At this point the team had spent considerable amounts of time optimising the previous design, with little time left to build and test the new concept. These observations lead to the key insight: **(KF 9) ChatGPT experiences and is limited by design fixation**, both regarding not wanting to abandon a working concept and specific details provided in prompts (e.g. the list of the available materials curated at the start of the hackathon). When planning the study and testing different initial prompts, it became clear that the LLM often fixated on specific details in the prompts. For example, when providing the LLM with available manufacturing capabilities available to the participants it would often fixate on the last machine on the list. If that happened to be a 3D printer, the LLM would suggest 3D printing each prototype going forward. Likewise, if the last machine were a laser cutter, it would suggest laser-cutting prototypes. This necessitated the general description of "fab lab/maker space" in the prompt instead of listing all manufacturing capabilities. Notably, the LLM asked participants to curate a list of available materials, but in never asked what manufacturing capabilities were available to them.

## 5.5 Recommendations for using current LLM for engineering design

This study intentionally utilized the LLM in an extreme capacity as the sole decision-maker in an engineering design process, a scenario that cannot be advocated for practical applications. The purpose was rather to objectively benchmark the strengths and weaknesses of current LLMs against human capabilities. Based on the observations and insights from the participants, it is clear that the results of using it could have been drastically improved through minimal critical thinking by them. Based on the findings of our study, we offer the following recommendations for effectively integrating current LLMs into the design process:

1. Leverage the LLM for ideation: Utilize its capability to generate a broad spectrum of concepts during the ideation phase to enhance creativity and explore a wider range of initial ideas.

2. Ensure human decision-making oversight: Incorporate human judgment to critically evaluate and select among the concepts generated by the LLM, especially to counteract the LLM's inclination to abandon promising ideas prematurely.

3. Implement iterative feedback loops: Continuous interaction between the LLM and human participants allows for the refinement of ideas and suggestions and prevents the addition of unnecessary complexity.

4. Create a custom GPT with templates and structured prompts: Streamline communication with the LLM through tailored prompts and templates, making instructions clearer and easier to follow.

5. Explicitly prompt the LLM to consider alternatives: Encourage the exploration of diverse solutions by directly asking the LLM to avoid design fixation and consider various design options.

6. Assign specific tasks at the subsystem level: Direct the LLM to focus on detailed explanations and solutions for specific project parts to enhance clarity and avoid vague or unhelpful responses.

## 5.6 Limitations

The study is limited by investigating the distinctive setting of a hackathon, with its specific characterisations that may not fully represent the larger range of design and prototyping processes found in professional and educational settings. However, hackathons have been shown to mirror key aspects of design (Ege, Goudswaard, et al., 2024), to be appropriate for studying design research (Goudswaard et al., 2022) and characterising and comparing design practices (Ege, Goudswaard, et al., 2024; Ege et al., 2023). Nonetheless, the insights gained are contextualized within this distinctive setting, and extrapolating these results to other design contexts should be done with caution. This study also focused exclusively on the prototyping towards a single task—a NERF firing device—thereby constraining the breadth of our insights into how the LLM might perform with different tasks. This specificity potentially limits the transferability of our findings to other cases. The replicability of the study may be compromised by the evolving nature of the LLM used. Results obtained on October 18th-19th, 2023, might not be replicated in subsequent uses due to updates and changes to the model.

Team 6 (ChatGPT) had one year more experience than control teams, a factor that could potentially influence outcomes. Choosing slightly more experienced participants for the team was a measure taken to ensure they had the abilities necessary to use all the available equipment and skills to perform the instructions of the LLM. Measures were taken to balance the experience gap, as the team was instructed to act objectively and not to contribute ideas or insights, thus neutralizing any differences. Further, participants' backgrounds in mechanical engineering may have inadvertently influenced the nature of the prompts given to the LLM, potentially introducing a bias. For instance, inquiries about the fit tolerances between parts may not typify questions from a novice, suggesting some experienced-based bias in participants' prompts.

The reliance on participants' objectivity is a further limitation. The LLM is shown to offer reasonable ideas and directions, but the physical realisation of these concepts depends on human intervention. This could introduce variability in decision-making and problem-solving, potentially distorting or undermining the findings, depending on the level of human engagement. The LLM's inability to effectively express design intent, introducing ambiguity and vagueness, might skew results due to interpretation from participants. Participants attempted to mitigate this by by seeking clarification from the LLM throughout the hackathon, yet this interactive process may itself influence the outcomes.

It is crucial to note that LLMs and their enabling technologies are evolving swiftly. The capabilities of current models are advancing, potentially making the recommendations and limitations described in this paper transient. The study reflects the state of LLMs as of the time of the research, and it is anticipated that issues such as the LLM's short attention span and logical reasoning will improve as the technology progresses.

## 5.7 Further work

The authors recommend further investigation of the recommendations presented in this paper for design-related tasks. Although the limitations placed on the participants were necessary to elicit the strengths and weaknesses of the LLM itself, they are not realistic in real-world design situations. This suggests that it would be interesting to investigate the LLM as a team member rather than the team lead role it was given in this study. Further, investigations of AI-generated visual outputs in conjunction with written building instructions could provide insights into effective cooperation strategies between designers and AI. This could be extended by incorporating parametric designs in code for AI-generated CAD to increase usability and lead to intuitive interactions, ultimately allowing for well-functioning human-AI cooperation for engineering design.

# 6 Conclusion

This study has compared the design practices and performance of a large language model, specifically ChatGPT 4.0, against 5th-year engineering students in a prototyping hackathon. It provides nine key findings, such as showing that the LLM had similar prototyping practices to human participants and proved competitive against them by finishing second among six teams. The LLM successfully provided building instructions to realise a physical, functional prototype and solved concrete and physical obstacles along the way. The concept generation capabilities of the LLM were particularly good. Among the limitations of the LLM is that it prematurely gave up on concepts when meeting what the authors perceived as minor difficulties and added unnecessary complexity to some of the designs. Communication between the LLM and participants was also challenging, as it often gave vague, too general or unclear descriptions and had trouble maintaining continuity and relevance in answers due to forgetting previously given information. The LLM also experienced design fixation by continuing iterations of one concept even though returns diminished instead of pivoting and trying alternative solutions. Based on these findings, we propose six recommendations for implementing current LLMs like ChatGPT in the design process, including leveraging it for ideation, ensuring human decision-making oversight, implementing iterative feedback loops, explicitly prompting it to consider alternatives and assigning the LLM specific tasks at a subsystem level.

## Competing interests statement

The authors declare none.

## Data availability statement

The data that support the findings of this study are openly available in Zenodo.org at http://doi.org/10.5281/zenodo.10495102, reference number 10495102, and described in the following data article: Ege, Øvrebø, Stubberud, Berg, Elverum, et al., 2024.

## Acknowledgements and funding

The authors would like to thank the TrollLabs Open hackathon participants for participating in this study.

The Research Council of Norway supports this research through its industrial PhD funding scheme (grant number 321386).

While preparing this work, the authors used ChatGPT by OpenAI to improve the language and readability of the manuscript. After using this tool, the authors reviewed and edited the content as needed and take full responsibility for the content of the publication.